\documentclass[preprint,prc,aps,tightenlines,floats,showpacs,showkeys,superscriptaddress,nofootinbib]{revtex4-1}
\usepackage{amsmath}
\usepackage{amssymb}
\usepackage[dvips]{graphicx,color}
\usepackage{dcolumn}
\usepackage{epsfig}
\usepackage{bm}

\usepackage{ulem}
\usepackage{color}
\usepackage{lineno}

\begin{document}

\title{
Isospin decomposition of $\gamma N \to N^*$ transitions within a dynamical coupled-channels model}
\author{H. Kamano}
\email{kamano@post.kek.jp}
\affiliation{KEK Theory Center, Institute of Particle and Nuclear Studies (IPNS), 
High Energy Accelerator Research Organization (KEK), Tsukuba, Ibaraki 305-0801, Japan}
\affiliation{J-PARC Branch, KEK Theory Center, IPNS, KEK, Tokai, Ibaraki 319-1106, Japan}
\author{S. X. Nakamura}
\affiliation{Department of Physics, Osaka University, Toyonaka, Osaka 560-0043, Japan}
\author{T.-S. H. Lee}
\affiliation{Physics Division, Argonne National Laboratory, Argonne, Illinois 60439, USA}
\author{T. Sato}
\affiliation{Department of Physics, Osaka University, Toyonaka, Osaka 560-0043, Japan}

\begin{abstract}
By extending the dynamical coupled-channels analysis performed in our previous work
[Phys.~Rev.~C {\bf 88}, 035209 (2013)] 
to include the available data of  photoproduction of $\pi$ meson off the neutron,
the transition amplitudes for the photo-excitation of the neutron to nucleon resonances, 
$\gamma n \to  N^*$, at the resonance pole positions are determined.  
The combined fits to the data for both the proton- and neutron-target reactions also
revise our results for the resonance pole positions 
and the $\gamma p \to N^*$ transition amplitudes.
Our results allow an isospin decomposition of the $\gamma N \to N^*$ transition amplitudes 
for the isospin $I=1/2$ $N^*$ resonances,
which is necessary for testing  hadron structure models
and gives crucial inputs for constructing models of 
neutrino-induced reactions in the nucleon resonance region.
\end{abstract}

\maketitle

\section{introduction}
\label{sec:intro}

Extracting the parameters, such as masses, widths, and form factors, associated with 
the $N$ and $\Delta$ resonances (collectively referred to as $N^*$) from 
the data of meson-production reactions 
is an important task for understanding Quantum Chromodynamics (QCD) in the nonperturbative domain.
It is usually done by performing a comprehensive partial-wave analysis of 
$\pi N$ and $\gamma N$ reactions with various final states including
$\pi N$, $\pi\pi N$, $\eta N$, $K\Lambda$, $K\Sigma$, and $\omega N$ etc.
within the on-shell $K$-matrix approaches~\cite{said,boga,giessen,pitt-anl} and the 
dynamical-model 
approaches~\cite{knls13,juelich,dmt}.
Here, the unitarity is a key essential element in accomplishing 
such a comprehensive analysis and making a reliable extraction of 
resonance parameters, because it not only ensures the conservation of 
probability in multichannel reaction processes 
but also properly defines the analytic structure of 
the scattering amplitudes in the complex-energy plane. 
The latter is crucial for 
extracting from the data correctly the 
resonance parameters defined at the poles of scattering amplitudes.

Among the $N^*$ resonance parameters, transition amplitudes for the photo- and electro-excitation 
of the nucleon to $N^*$ resonances, $\gamma^{(*)} N \to N^*$,
have been a particular focus of interest in the $N^*$ spectroscopy 
because of its crucial role in understanding the electromagnetic properties
as well as the quark-gluon substructure of $N^*$ resonances (see, e.g., Ref.~\cite{bl04}).
In fact, the dynamical-model studies~\cite{jlss07,jlmss08,jklmss09,ssl10} of photon- and 
electron-induced meson-production reactions off the proton target
revealed the large meson-cloud effect on the transition amplitudes at low $Q^2$
whereas it decreases when $Q^2$ is increased, implying that
the (constituent) quark-gluon core of $N^*$
surrounded by dense meson clouds at long distance scales
gradually emerges at shorter distance scales.
In this regard, new extensive measurements of $ep\to e'X$ with $X=\pi N, \pi \pi N, KY$ 
are planned at CLAS12~\cite{clas12-proposals}, aiming at
precise determinations of the $\gamma^{(*)} p \to N^*$ transition amplitudes in the $Q^2$ region
where such a transition to the ``core-dominated'' region is expected to occur.

However, for the $I=1/2$ $N^*$ resonances, the $\gamma^{(*)} N \to N^*$ transition amplitudes
for both the proton ($N=p$) and neutron ($N=n$)
are necessary to uniquely determine the isospin structure of the photo- and electro-excitation 
couplings.
The information on such isospin-decomposed transition amplitudes 
are also important for investigating the 
neutrino-induced reactions in the $N^*$ resonance
region (see, e.g., Ref.~\cite{nks15}) because those are the basic ingredients for constructing 
the vector-current matrix elements associated with the weak interactions.
The analysis of the photo- and electro-production reactions
on both the proton and neutron targets is required
to determine the isospin structure of 
the $\gamma^{(*)} N \to N^*$ transition amplitudes for the $I=1/2$ $N^*$ resonances\footnote{
For the $I = 3/2$ $N^*$ resonances,
the $\gamma^{(*)} p \to N^*$ and $\gamma^{(*)} n \to N^*$ transition amplitudes
are the same, and thus in principle
only the data for either the proton- or neutron-target reaction is required.
However, it is still highly desirable to have the data for both reactions
so that one can have more constraints on the transition amplitudes for 
$I = 3/2$ $N^*$ resonances as well.}, 
and such combined analyses have been done so far by several analysis groups~\cite{hua11,ani13}
within the coupled-channels framework.

In Ref.~\cite{knls13}, we reported on the $N^*$ parameters extracted within
the dynamical coupled-channels (DCC) model developed in Ref.~\cite{msl07}.  
Our approach is to fit the  meson production data and to also search for the resonance poles
by solving the following coupled integral equations for the partial-wave amplitudes,
$T_{\beta,\alpha}(p_\beta,p_\alpha;W)$, 
which are specified by
the total angular momentum $J$, parity $P$, and total isospin $I$ (these indices are suppressed
in the following equations),
\begin{eqnarray}
T_{\beta,\alpha}(p_\beta,p_\alpha;W)= V_{\beta,\alpha}(p_\beta,p_\alpha; W)
+ \sum_{\delta}
 \int p^{2}d p  V_{\beta,\delta}(p_\beta, p;W )
G_{\delta}(p; W)
T_{\delta,\alpha}( p  ,p_\alpha;W)  \,,
\label{eq:teq}
\end{eqnarray}
with
\begin{eqnarray}
V_{\beta,\alpha}(p_\beta,p_\alpha; W)= 
v_{\beta,\alpha}(p_\beta,p_\alpha)+
Z_{\beta,\alpha}(p_\beta,p_\alpha;W)+
\sum_{N^*_n}\frac{\Gamma_{\beta,N^*_n}(p_\beta)
 \Gamma_{N^*_n,\alpha}(p_\alpha)} {W-M^0_{N^*_n}} .
\label{eq:veq}
\end{eqnarray}
Here, the subscripts 
$\alpha,\beta,\delta = \pi N, \eta N, K\Lambda, K\Sigma, \pi \Delta, \rho N, \sigma N$
represent reaction channels considered,
for which the $\pi \Delta$, $\rho N$, and $\sigma N$ channels are 
the resonant components of the three-body $\pi\pi N$ channel.
[Indices for the orbital angular momentum ($L$) and total spin ($S$) of each reaction channel are
suppressed.]
$G_\delta (p,W)$ is the Green's function of channel $\delta$; 
$M^0_{N^*_n}$ is the mass of the $n$th bare excited nucleon state $N^*_n$ in a given partial wave; 
the hadron-exchange potential $v_{\beta,\alpha}$ is derived from the effective Lagrangians
by making use of the unitary transformation method\footnote{The potential derived with 
the unitary transformation method becomes energy independent.
The off-shell behavior of the potential is also defined within this method.}~\cite{sko,sljpg};
the energy-dependent $Z_{\beta,\alpha}(p_\beta,p_\alpha;W)$ term~\cite{msl07} is the effective
one-particle-exchange potential that is derived with the projection operator method~\cite{feshbach}
and produces the three-body $\pi \pi N$ cut; 
the vertex interaction $\Gamma_{\alpha,N^*_n}$ defines the $N^*_n \to \alpha$ decay
(note $\Gamma_{\alpha,N^*_n} =\Gamma^\dag_{N^*_n,\alpha}$). 
Similar approach is also taken in Ref.~\cite{juelich}.
The differences between our approach and the other coupled-channels analyses 
\cite{boga,giessen,pitt-anl}, which only consider the on-shell
matrix elements of $T_{\alpha,\beta}(p_\alpha,p_\beta;W)$, 
have been discussed in detail in Refs.~\cite{msl07,knls13,knls14}. 
Here we only mention that these models can be qualitatively obtained from 
Eq.~(\ref{eq:teq}) by keeping only the on-shell part of the propagator $G_\delta(p;W)$, 
and Refs.~\cite{boga,pitt-anl} further
replace the hadron-exchange interaction $V_{\beta,\alpha}$ by phenomenological forms
such as the polynomials of on-shell momenta.
If the data are \textit{complete} (as explained, e.g., in Ref.~\cite{shkl11})
and the high accuracy fits can be achieved, all approaches are acceptable for extracting
the resonance pole positions. 
However, more investigations are needed to examine under what ideal conditions 
all approaches should give the same resonance parameters defined at the poles of the scattering
amplitudes. 
Furthermore, it is practically impossible to get \textit{complete} data. 
Thus it is essential to impose  theoretical constraints on both the determinations
of the partial-wave amplitudes and the extractions of $N^*$ parameters.  
This is accomplished in our approach by implementing the well-established 
hadron-exchange mechanisms, as defined by $v_{\beta,\alpha}$ in Eq.~(\ref{eq:veq}), in the fits. 
This also allows us to provide interpretations of the structure of
the extracted $N^*$ resonances, such as the meson-cloud effects on the $\gamma^{(*)} N \to N^*$ 
transitions.

For the calculations of the $\gamma N$ reaction amplitudes, 
we use the so-called helicity-$LSJ$ mixed-representation~\cite{msl07},
in which the initial $\gamma N$ state is specified
by their helicities, $\lambda_\gamma$ and $\lambda_N$, while the final meson-baryon
state is specified by $L$, $S$, $J$ and $I$ as in Eq.~(\ref{eq:teq}),
\begin{eqnarray}
T_{\beta,\gamma N(\lambda)}(p_\beta,q;W)&=& 
V_{\beta,\gamma N(\lambda)}(p_\beta,q; W)
\nonumber\\
&&\quad
+ \sum_{\delta}
 \int p^{2}d p T_{\beta,\delta}( p_\beta  ,p;W) 
G_{\delta}(p; W)
V_{\delta,\gamma N(\lambda)}(p,q; W) \,,
\label{eq:teq2}
\end{eqnarray}
\begin{eqnarray}
V_{\beta,\gamma N(\lambda)}(p_\beta,q; W)= 
v_{\beta,\gamma N(\lambda)}(p_\beta,q)+
\sum_{N^*_n}\frac{\Gamma_{\beta,N^*_n}(p_\beta) \Gamma_{N^*_n,\gamma N(\lambda)}(q)} {W-M^0_{N^*_n}} \,,
\label{eq:veq2}
\end{eqnarray}
where $\lambda = \lambda_\gamma - \lambda_N$.
Here we note that the summation in Eq.~(\ref{eq:teq2}) runs over only hadronic meson-baryon channels.
We take $\gamma N$ channel perturbatively since the electromagnetic interactions are
much smaller than the strong ones and their effect on the resonance parameters are expected
to be just the order of isospin breaking.
The potential $v_{\beta,\gamma N(\lambda)}(p_\beta,q)$ is again derived from the effective Lagrangians
by making use of the unitary transformation method.
On the other hand, in Ref.~\cite{juelich} 
the potential for electromagnetic interaction is simply parametrized with polynomials.

In our previous work performed in Ref.~\cite{knls13},
we analyzed the available data of $\pi p, \gamma p \to \pi N, \eta N, K\Lambda, K\Sigma$
reactions in the region of $W\lesssim$ 2.1 GeV.
Then 24 physical $N^*$ resonances, which are defined at the  poles of the scattering amplitudes
in the complex energy plane, were successfully extracted. 
Their properties, including the $\gamma p \to N^*$ transition amplitudes,  
were also extracted by evaluating the residues of the scattering amplitudes at the resonance poles.

As a first step towards understanding the isospin structure of 
the photo- and electro-excitation of the nucleon to the $I=1/2$ $N^*$
resonances, in this work we extend our previous DCC analysis~\cite{knls13}
by further including the available data of $\pi$ photoproductions off the neutron
and making a combined analysis of meson production reactions off the proton and neutron targets.
We then present the extracted  $\gamma n \to N^*$ transition amplitudes,
together with the improved results for resonance pole masses, $\pi N$ elastic residues,
and the $\gamma p \to N^*$ transition amplitudes.  
In this work, we focus on studying the transition amplitudes at the photon point, $Q^2=0$.

Our procedures of resonance extraction have been given 
in detail in Refs.~\cite{ssl09,sjklms10,ssl10,knls13,knls15}.
We therefore will only recall in Sec.~\ref{sec:formulas} the formulae that 
are needed for presenting the parameters of the extracted
$\gamma N\to N^*$ transitions. 
In Sec.~\ref{sec:fit}, we present our fits to the data.
The extracted $\gamma n\to N^*$ transition amplitudes are presented 
in Sec.~\ref{sec:respar}, 
along with the revised values of resonance pole positions and 
the $\gamma p\to N^*$ transition amplitudes presented in Ref.~\cite{knls13}.
In Sec.~\ref{sec:summary}, we give a summary and discussions 
on the necessary future works.

\section{Formulas for the  $\gamma N \to N^*$ transition amplitudes}
\label{sec:formulas}

To define the $\gamma N \to N^*$ transition amplitudes,
we recall here some formula that can be derived~\cite{knls13,msl07,knls15}
from Eqs.~(\ref{eq:teq}) and~(\ref{eq:teq2}) within the considered
dynamical coupled-channels model.
The on-shell $S$ matrix elements of 
the meson-baryon reactions, $MB \to M'B'$, 
in the center-of-mass system are given for each partial wave by
\begin{equation}
S_{M'B',MB}(W) = \delta_{M'B',MB} + 2iF_{M'B',MB}(W).
\label{eq:S}
\end{equation}
Here $W$ is the total scattering energy
 and 
we have suppressed indices for the angular momenta, parity, and isospin quantum numbers
associated with the channels $MB$ and $M'B'$.
The on-shell scattering amplitudes $F_{M'B',MB}(W)$ are related to
the $T$ matrix elements given by Eq.~(\ref{eq:teq}) as follows:
\begin{equation}
F_{M'B',MB}(W) = -
[\rho_{M'B'}(k_{M'B'}^{\text{on}};W)]^{1/2} 
T_{M'B',MB}(k_{M'B'}^{\text{on}}, k_{MB}^{\text{on}}; W) 
[\rho_{MB}(k_{MB}^{\text{on}};W)]^{1/2},
\label{eq:F-T}
\end{equation}
with
\begin{equation}
\rho_{MB}(k_{MB};W) = \pi \frac{k_{MB} E_M(k_{MB}) E_B(k_{MB})}{W},
\label{eq:rho}
\end{equation}
where $E_\alpha(k_{\alpha}) = \sqrt{m_\alpha^2+k_\alpha^2}$ is the energy of a particle $\alpha$
with  mass $m_\alpha$ and  three-momentum $\vec k_\alpha$ ($k_\alpha \equiv |\vec k_{\alpha}|$).
For a given $W$, which can be complex, the on-shell momentum for the channel 
$MB$, $k_{MB}^{\mathrm{on}}$, is defined by $W = E_M (k_{MB}^{\mathrm{on}}) + E_B (k_{MB}^{\mathrm{on}})$.

As the energy $W$ approaches to a pole position $M_R$ in the complex $W$ plane,
the scattering amplitudes take the following form,
\begin{equation}
F_{M'B',MB}(W\to M_R) = -\frac{R_{M'B',MB}}{W-M_R} + B_{M'B',MB} ,
\label{eq:F-pole}
\end{equation}
where $R_{M'B',MB}$ is the residue of $F_{M'B',MB}(W)$ at the resonance pole $M_R$ and
$B_{M'B',MB}$ is the ``background'' contribution. 
Both $R_{M'B',MB}$ and $B_{M'B',MB}$ are constant and in general complex.
The pole position ($M_R$) and the residue ($R_{M'B',MB}$) are fundamental quantities
that characterize the extracted resonance.

The residues $R_{M'B',MB}$ defined in Eq.~(\ref{eq:F-pole}) can be calculated by using the 
definition
\begin{equation}
R_{M'B',MB} = \frac{1}{2\pi i}\oint_{C_{M_R}} dW [-F_{M'B',MB}(W)],
\label{eq:residue_def}
\end{equation}
where $C_{M_R}$ is an appropriate closed-path in the neighborhood of the point $W = M_R$,
circling $W = M_R$ in a counterclockwise manner.
It can be shown that for partial waves with one or more bare states
(this is the case of our current model for all partial waves),
$R_{M'B',MB}$ can also be calculated~\cite{ssl09,ssl10,knls13} with
\begin{equation}
R_{M'B',MB} = 
[\rho_{M'B'}(k_{M'B'}^{\text{on}};M_R)]^{1/2} 
\bar \Gamma^R_{M'B'}(k_{M'B'}^{\text{on}};M_R)
\bar \Gamma^R_{MB}(k_{MB}^{\text{on}};M_R)
[\rho_{MB}(k_{MB}^{\text{on}};M_R)]^{1/2}.
\label{eq:residue2}
\end{equation}
Here $\bar \Gamma^R_{MB}(k_{MB}^{\text{on}};M_R)$ is the 
(renormalized) dressed $MB \to N^*$ vertex that contains the meson cloud 
arising from the coupling to the meson-baryon continuum states.

\begin{figure}
\includegraphics[clip,width=0.6\textwidth]{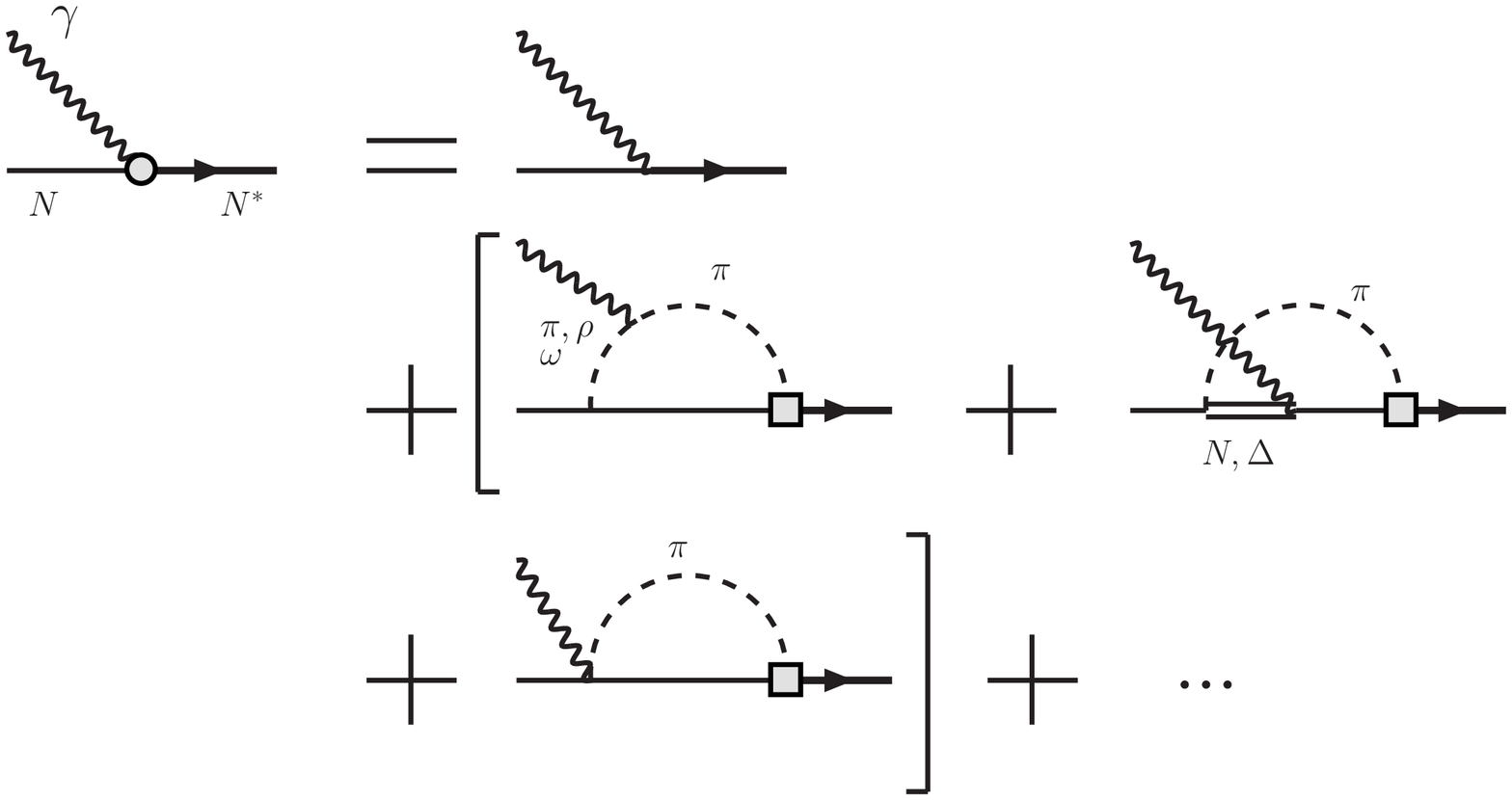}
\caption{\label{fig:dressed-v}Graphical representation of dressed $\gamma N \rightarrow N^*$
transition vertex.}
\end{figure}

As for the $\gamma N \to M'B'$ reactions, the on-shell scattering amplitudes 
are given in the helicity-$LSJ$ mixed-representation by
\begin{equation}
F_{M'B',\gamma N(\lambda)}(W) = 
-[\rho_{M'B'}(k_{M'B'}^{\text{on}};W)]^{1/2} 
T_{M'B',\gamma N(\lambda)}(k_{M'B'}^{\text{on}}, k_{\gamma N}^{\text{on}}; W) 
[\rho_{\gamma N}(k_{\gamma N}^{\text{on}};W)]^{1/2},
\label{eq:F-T-ele}
\end{equation}
where $k_{\gamma N}^{\text{on}}$ is given by 
$W = k_{\gamma N}^{\text{on}} + E_N(k_{\gamma N}^{\text{on}})$.
The amplitude $F_{M'B',\gamma N(\lambda)}(W)$ also has a form close to 
the resonance pole position $M_R$,
\begin{equation}
F_{M'B',\gamma N(\lambda)}(W\to M_R) = -\frac{R_{M'B',\gamma N(\lambda)}}{W-M_R} + B_{M'B',\gamma N(\lambda)} .
\label{eq:F-pole-em}
\end{equation}
Then, the residue at the resonance pole $M_R$, $R_{M'B',\gamma N(\lambda)}$, can be calculated
using the same formula~(\ref{eq:residue_def}) as the $MB\to M'B'$ cases, or can be calculated 
using the similar formula to Eq.~(\ref{eq:residue2}), 
\begin{equation}
R_{M'B',\gamma N(\lambda)} = 
[\rho_{M'B'}(k_{M'B'}^{\text{on}};M_R)]^{1/2} 
\bar \Gamma^R_{M'B'}(k_{M'B'}^{\text{on}};M_R)
\bar \Gamma^R_{\gamma N(\lambda)}(k_{\gamma N}^{\text{on}};M_R)
[\rho_{\gamma N}(k_{\gamma N}^{\text{on}};M_R)]^{1/2},
\label{eq:residue2g}
\end{equation}
where the quantity $\bar \Gamma^R_{\gamma N(\lambda)}(k_{\gamma N}^{\text{on}};M_R)$  
is the dressed $\gamma N \to N^*$ vertex, as illustrated in Fig.~\ref{fig:dressed-v},
multiplied by an appropriate wave-function renormalization factor.
The details on how to compute $\bar \Gamma^R_{MB}(k_{MB}^{\text{on}};M_R)$ and
$\bar \Gamma^R_{\gamma N(\lambda)}(k_{\gamma N}^{\text{on}};M_R)$ etc.
have been given in Refs.~\cite{ssl09,ssl10,knls13}.

With the renormalized vertex $\bar{\Gamma}^R_{\gamma N(\lambda)}(k_{\gamma N}^{\rm on};M_R)$,
the helicity amplitudes for the $\gamma N \to N^*$ transitions 
evaluated at the resonance pole position $M_R$, 
$A_{3/2}$ and $A_{1/2}$,
can be written as~\cite{ssl10}  
\begin{eqnarray}
A_{3/2}  &=&  C\times \bar{\Gamma}^R_{\gamma N(3/2)}(k_{\gamma N}^{\text{on}};M_R),
\label{eq:a32}
\\
A_{1/2}  &=&  C\times \bar{\Gamma}^R_{\gamma N(1/2)}(k_{\gamma N}^{\text{on}};M_R),
\label{eq:a12}
\end{eqnarray}
with
\begin{eqnarray}
C=\sqrt{\frac{E_N(k_{\gamma N}^{\text{on}})}{m_N }} \frac{1}{\sqrt{2K}} 
\sqrt{\frac{(2J^R+1)(2\pi)^32|k_{\gamma N}^{\text{on}}|}{4\pi}},
\label{eq:coef-c}
\end{eqnarray}
where $J^R$ is the spin of the resonance state and
$K  =  (M_R^2 - m_N^2)/(2M_R)$.

By fitting to the data, the model parameters included in the potentials~(\ref{eq:veq}) 
and~(\ref{eq:veq2}) are determined.
We then can compute the on-shell dressed $\gamma N \to N^*$ vertices at the resonance pole 
position $\bar{\Gamma}^R_{\gamma N(\lambda)}(k_{\gamma N}^{\text{on}};M_R)$
and determine the helicity amplitudes $A_{\lambda}$ by using 
Eqs.~(\ref{eq:a32})-(\ref{eq:coef-c}).
In practice, however, we use an alternative formula from Ref.~\cite{hel-phase}
to calculate the helicity amplitudes and
unambiguously fix their phases.
The formula is given with the residues
$R_{\pi N,\gamma N(\lambda)}$ and $R_{\pi N,\pi N}$ as follows:
\begin{eqnarray}
A_{3/2} & = & N \times R_{\pi N \gamma N(3/2)},
\\
A_{1/2} & = & N \times R_{\pi N \gamma N(1/2)},
\end{eqnarray}
\begin{equation}
N = a \times \sqrt{\frac{k^{\rm on}_{\pi N}}{K}
\frac{2\pi(2J^R+1)M_R}{ m_N R_{{\pi N, \pi N}}}} ,
\end{equation}
with $a=\sqrt{2/3}$ for $I$=3/2 $N^*$ and $a=-\sqrt{3}$ for $I$=1/2 $N^*$;
the phase is fixed so that $-\pi/2 \le \arg(N/a) \le \pi/2 $.

\section{Combined analysis of meson productions off the proton and neutron targets}
\label{sec:fit}

\begin{table}
\caption{\label{tab:data-chi2}
Observables and number of the data points considered in this coupled-channels analysis.
The data are taken from the database of the INS DAC Services~\cite{cns}.
}
\begin{ruledtabular}
\begin{tabular}{lcc}
Reactions &Observables & Number of data points\\
\hline
$\gamma \text{`}n\text{'} \to \pi^- p$ &$d\sigma/d\Omega$& 2305  \\
                       &$\Sigma$& 308  \\
                       &$T$& 94  \\
                       &$P$& 88  \\
$\gamma \text{`}n\text{'} \to \pi^0 n$ &$d\sigma/d\Omega$& 148 \\
                       &$\Sigma$& 216 \\
\end{tabular}
\end{ruledtabular}
\end{table}

In this work, we have performed combined fits to the data for 
both the proton- and neutron-target reactions.
The same database as used in our previous work~\cite{knls13} is employed
for the proton-target reactions,
$\pi p\to MB$ and $\gamma p \to MB$ with $MB = \pi N, \eta N, K\Lambda, K\Sigma$,
while for the neutron-target reactions we include
$\sim 3,200$ data points of $\gamma\, `n\textrm{'} \to \pi N$
as summarized in Table~\ref{tab:data-chi2} 
(where $\text{`}n\text{'}$ represents the neutron bound in the deuteron).

The fitting procedures are the same as explained in Ref.~\cite{knls13} and 
will not be repeated here.
The results of our fits to the data for the neutron-target reactions 
are presented in Sec.~\ref{subsec:n}.
As for the fits to the much more extensive data for the proton-target reactions,
however, we will only present a small sample of the selected results in Sec.~\ref{subsec:p}
to indicate the differences with our fits presented in Ref.~\cite{knls13}.
The values of determined model parameters are given in the supplemental material~\cite{suppl}.

\subsection{Fits to the data for the neutron-target reactions}
\label{subsec:n}

\begin{figure}
\includegraphics[clip,width=0.75\textwidth]{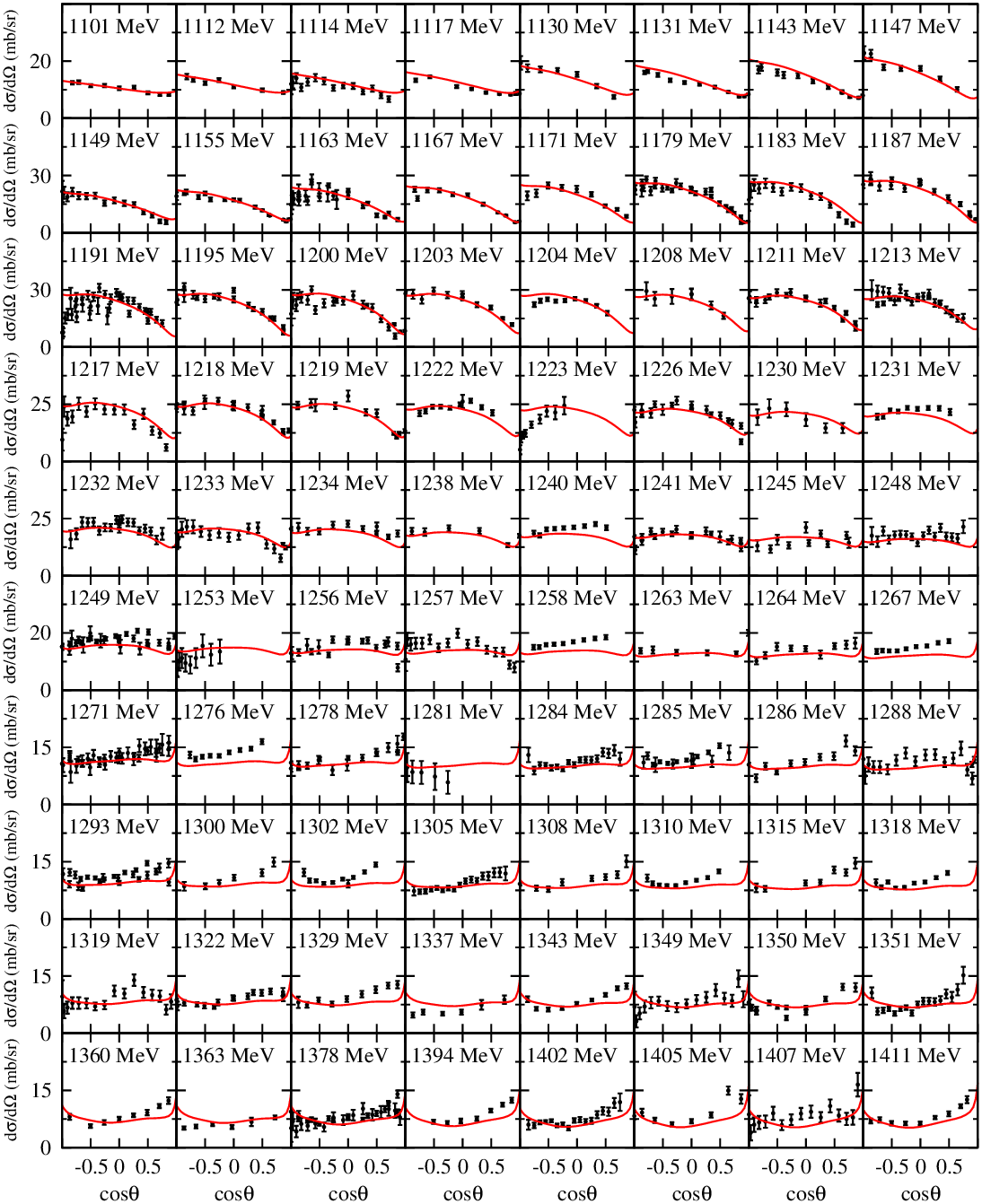}
\caption{\label{fig:gnpimp-dc-1} (Color online) Fitted result for $d\sigma/d\Omega$ of $\gamma\, `n\textrm{'} \to \pi^- p$.}
\end{figure}
 
\begin{figure}
\includegraphics[clip,width=0.75\textwidth]{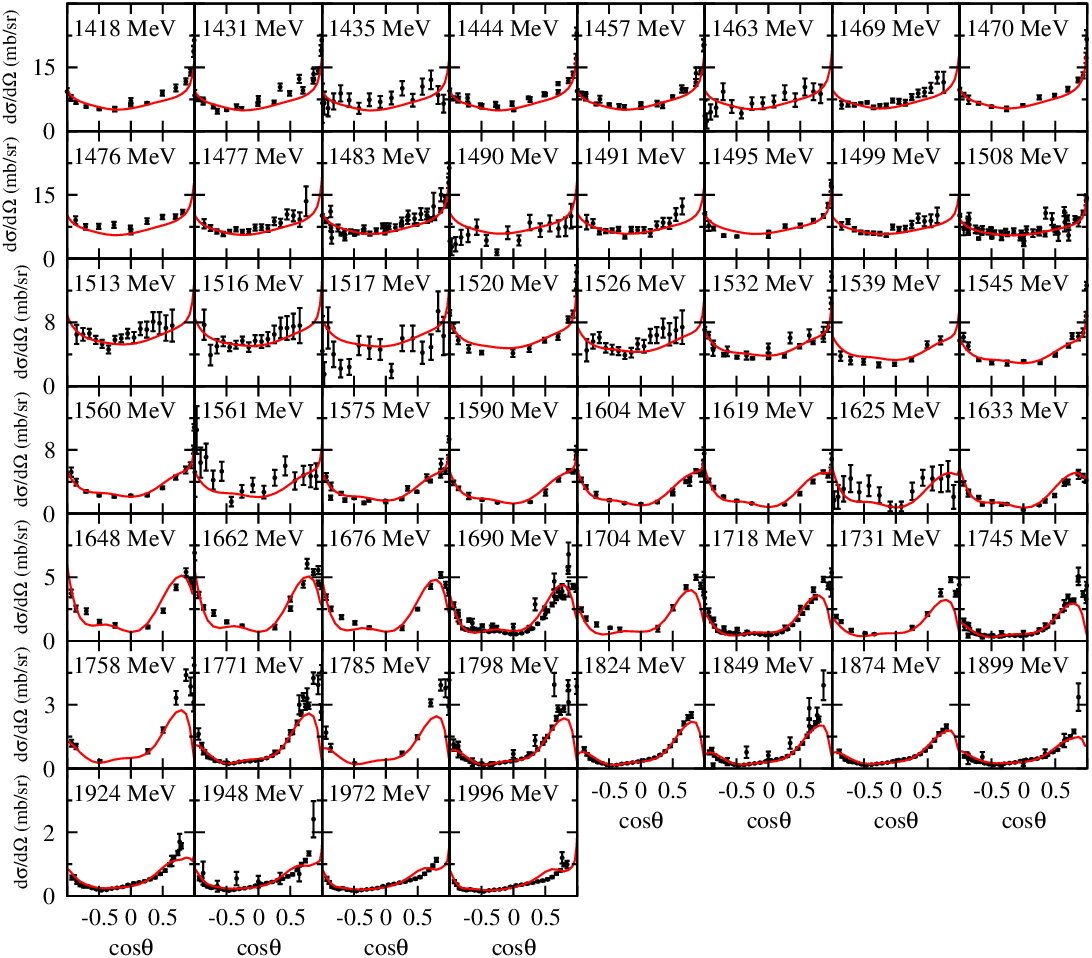}
\caption{\label{fig:gnpimp-dc-2} (Color online) Fitted result for $d\sigma/d\Omega$ of $\gamma\, `n\textrm{'} \to \pi^- p$ (continued).}
\end{figure}
 
\begin{figure}
\includegraphics[clip,width=0.75\textwidth]{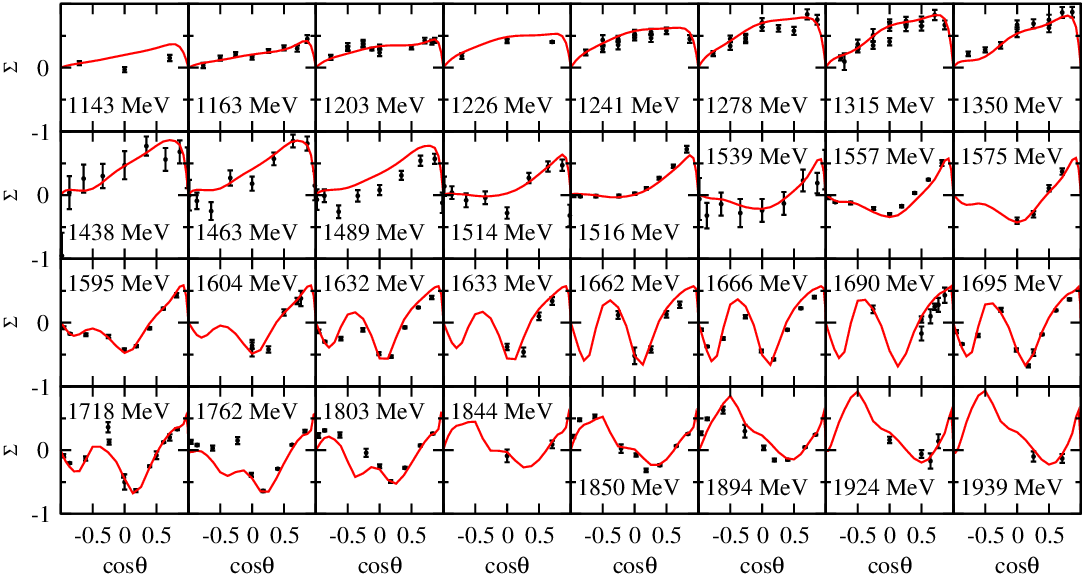}
\caption{\label{fig:gnpimp-s} (Color online) Fitted result for $\Sigma$ of $\gamma\, `n\textrm{'} \to \pi^- p$.}
\end{figure}
 
\begin{figure}
\includegraphics[clip,width=0.75\textwidth]{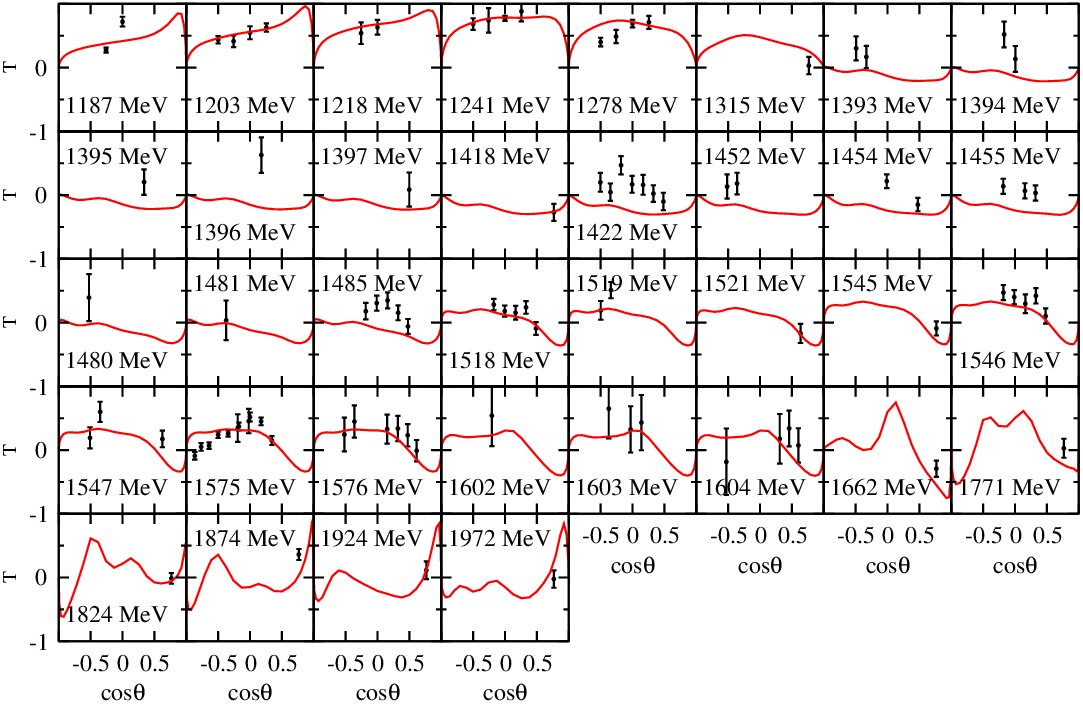}
\caption{\label{fig:gnpimp-t} (Color online) Fitted result for $T$ of $\gamma\, `n\textrm{'} \to \pi^- p$.}
\end{figure}
 
\begin{figure}
\includegraphics[clip,width=0.75\textwidth]{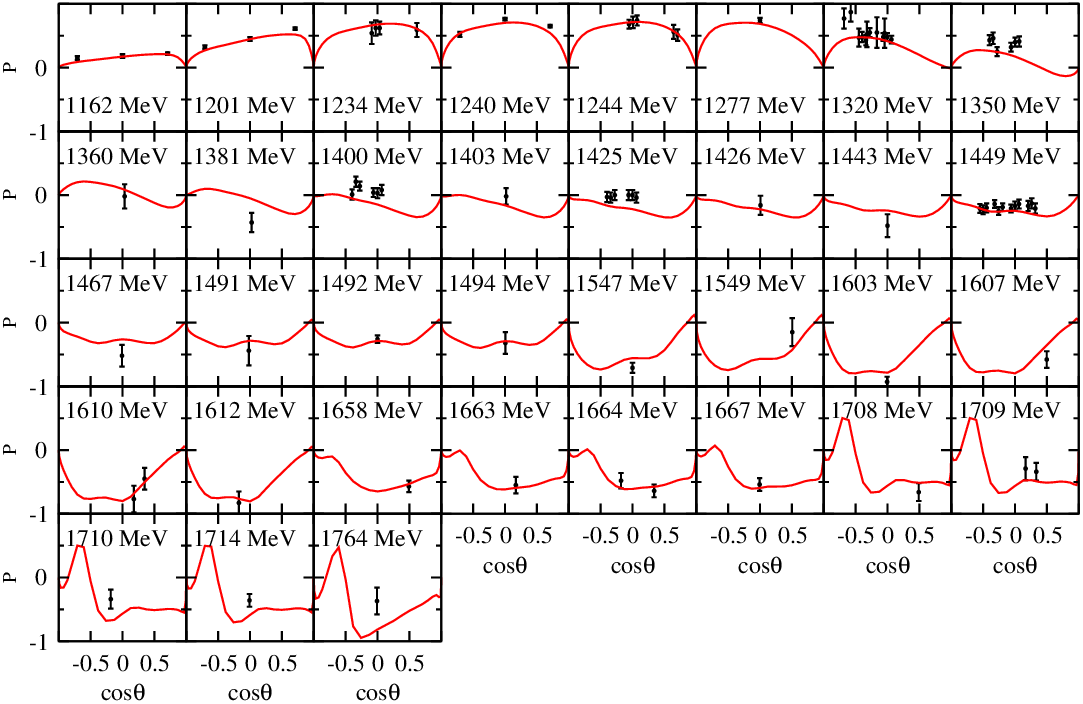}
\caption{\label{fig:gnpimp-p} (Color online) Fitted result for $P$ of $\gamma\, `n\textrm{'} \to \pi^- p$.}
\end{figure}

\begin{figure}
\includegraphics[clip,width=0.48\textwidth]{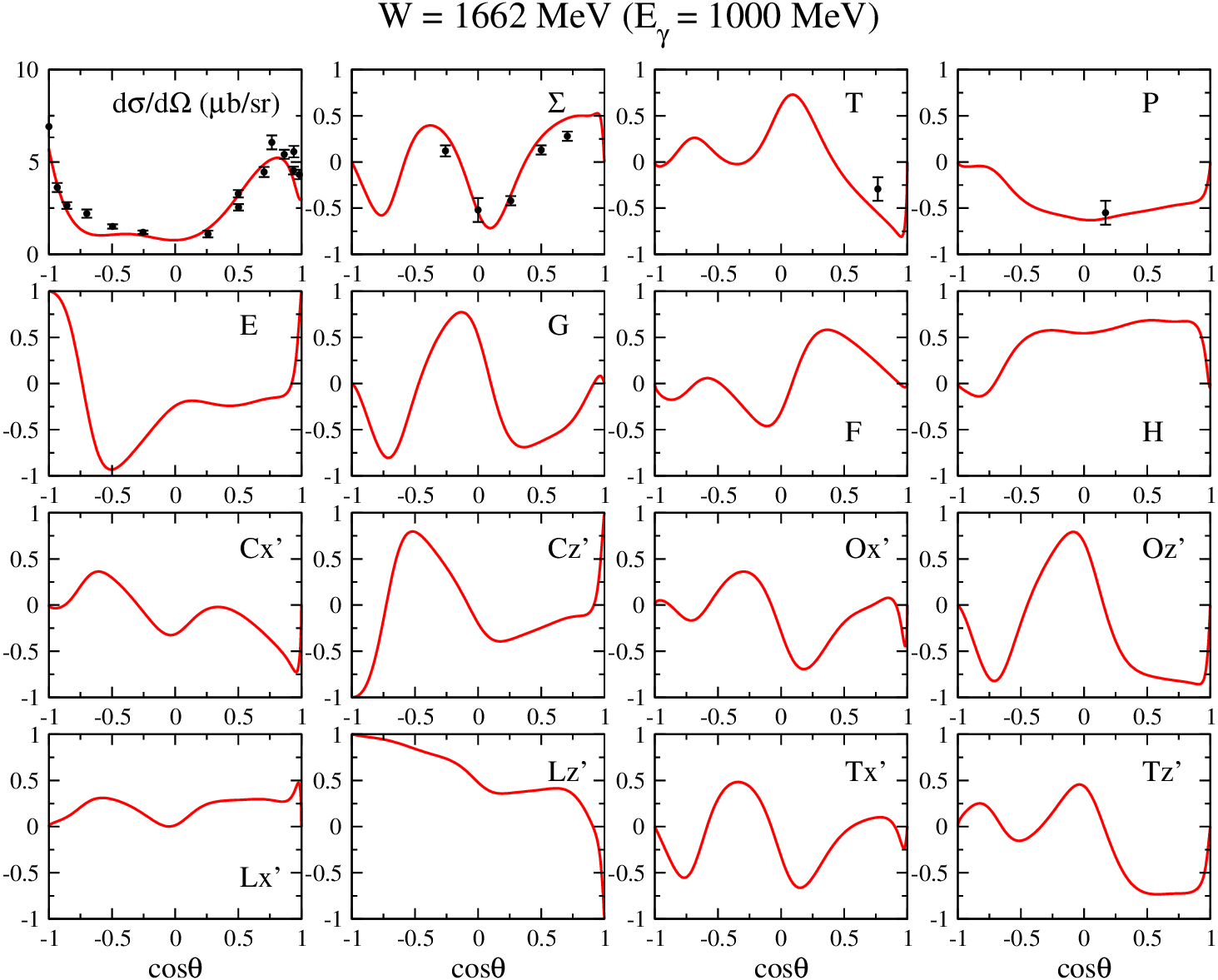}
\quad
\includegraphics[clip,width=0.48\textwidth]{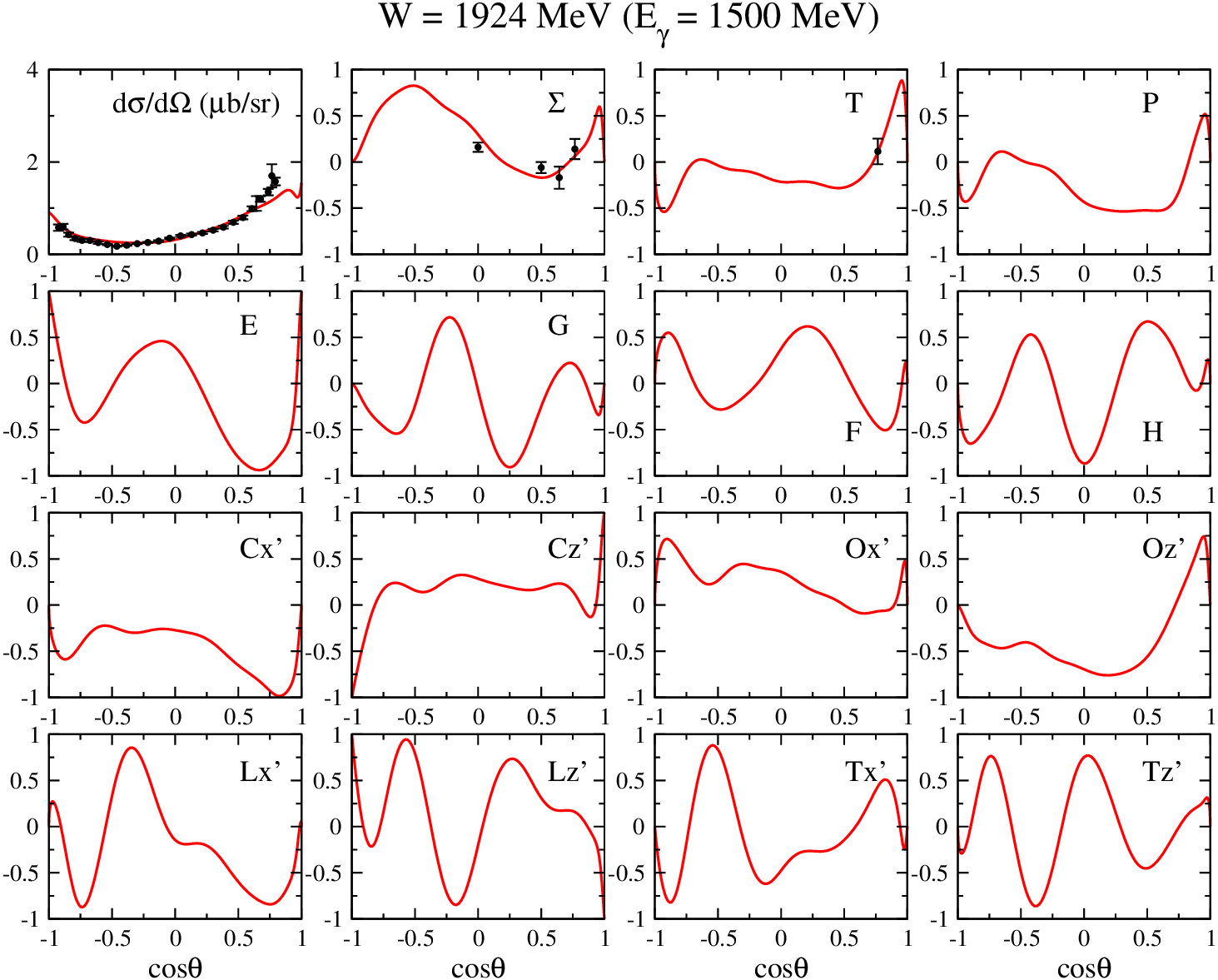}
\caption{\label{fig:all} (Color online) 
Differential cross section and all polarization observables for $\gamma\, `n\textrm{'} \to \pi^- p$.
The left panels [right panels] are the results at $W = 1662$ MeV ($E_\gamma = 1000$ MeV)
[$W = 1924$ MeV ($E_\gamma = 1500$ MeV)].
The results for the observables for which no data are presented 
are predictions from our current DCC model.
}
\end{figure}

The results of our fits to the $\gamma\, `n\textrm{'} \to \pi^- p$ data are presented
in Figs.~\ref{fig:gnpimp-dc-1} and~\ref{fig:gnpimp-dc-2} for the differential cross section ($d\sigma/d\Omega$),
Fig.~\ref{fig:gnpimp-s} for the photon asymmetry ($\Sigma$),
Fig.~\ref{fig:gnpimp-t} for the target polarization ($T$), and
Fig.~\ref{fig:gnpimp-p} for the recoil polarization ($P$).
Clearly, the results are reasonably good in the considered energy region up to $W = 2$ GeV.
However, it should be noted that
more data for polarization observables would be highly desirable
to determine $\gamma n \to N^*$ transition amplitudes.
In this regard, a number of new data for meson photoproductions using 
the polarized photon beam and/or the polarized deuteron target 
will be available soon from the electron and photon beam facilities 
such as JLab, ELSA, and MAMI.
The main purpose of these experimental efforts is to obtain precise 
data for the ``(over-)complete'' set of observables
for the $\gamma\, `n\textrm{'} \to MB$ reactions with $MB=\pi N, \eta N, K\Lambda, K\Sigma,\cdots$,
which, together with the  data of the reactions on the proton target, 
are crucial for obtaining definitive
information for the isospin structure of photo-excitation amplitudes of the nucleon.
For a reference, in Fig.~\ref{fig:all} we present the results of all 16 observables
for the $\gamma\, `n\textrm{'} \to \pi^- p$ reaction 
at $W = 1662$ MeV and $W = 1924$ MeV, which respectively correspond to 
$E_\gamma = 1$ GeV and $E_\gamma = 1.5$ GeV with $E_\gamma$ being the incoming 
photon energy in the Lab frame.
The calculated results of all observables for 
$\gamma n \to \pi^- p$ as well as $\gamma n \to \pi^0 n$ for $W \leq 2$ GeV 
are available upon request.

While the results of our fits to the $\gamma\, `n\textrm{'} \to \pi^- p$ data are reasonably 
good, we also see some deviations of our curves from the data points at several energies.
One reason for this would be due to an inconsistency between different data sets.
For example,  
there are  significant differences between the $d\sigma/d\Omega$ data at $W = 1240$ and 1241 MeV 
in magnitude for $\cos\theta > 0$,
even though they have just 1 MeV difference in $W$. These two data sets 
were extracted from the $\gamma d \to \pi^- pp$ reaction
by different experiment/analysis groups
in Ref.~\cite{briscoe12} and in Refs.~\cite{fujii71,benz73,fujii77}, respectively.
This kind of inconsistency  could arise from the differences in the methods employed for extracting
the information on the $\gamma\, `n\textrm{'}$ reactions from the $\gamma d$ reactions, 
e.g., the momentum cuts taken for the data selection and/or the initial/final state interactions
taken into account, etc.
The $\gamma\, `n\textrm{'} \to \pi N$ data are thus ``biased'' by the
employed analysis methods.
To avoid such inconsistency, ideally one should directly analyze 
the original $\gamma d$ reaction data.
 Nevertheless, as a first step toward
a reliable extraction of the $\gamma n \to N^*$ amplitudes, we follow the previous
analyses by using the available $\gamma\, `n\textrm{'} \to \pi N$ data in this work.
The model parameters obtained in this work are then used as
starting values for the $\gamma d$ analysis that will be presented elsewhere.

\begin{figure}
\includegraphics[clip,width=0.75\textwidth]{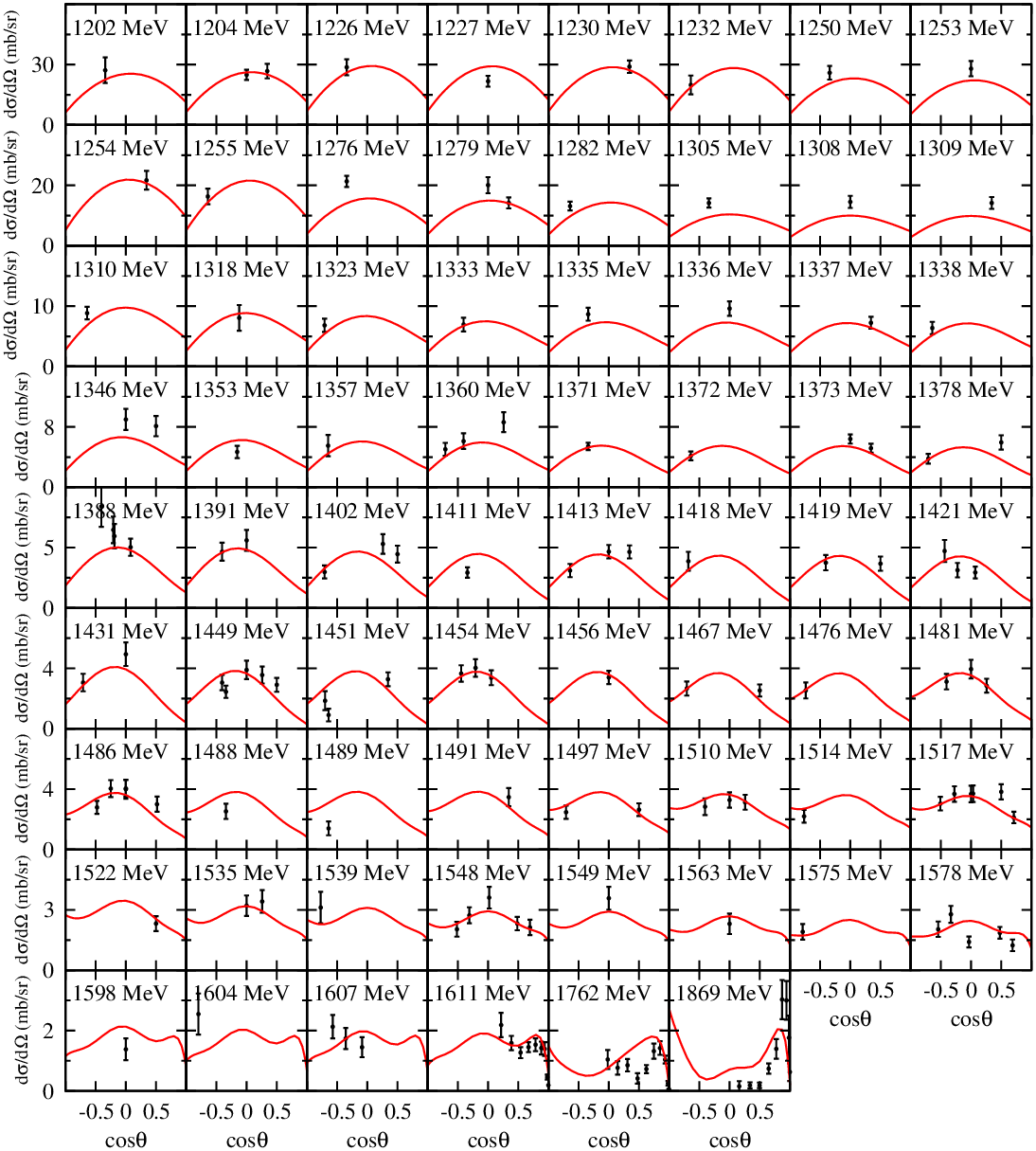}
\caption{\label{fig:gnpi0n-dc}(Color online) Fitted result for  $d\sigma/d\Omega$ of $\gamma\, `n\textrm{'} \to \pi^0 n$.}
\end{figure}

\begin{figure}
\includegraphics[clip,width=0.75\textwidth]{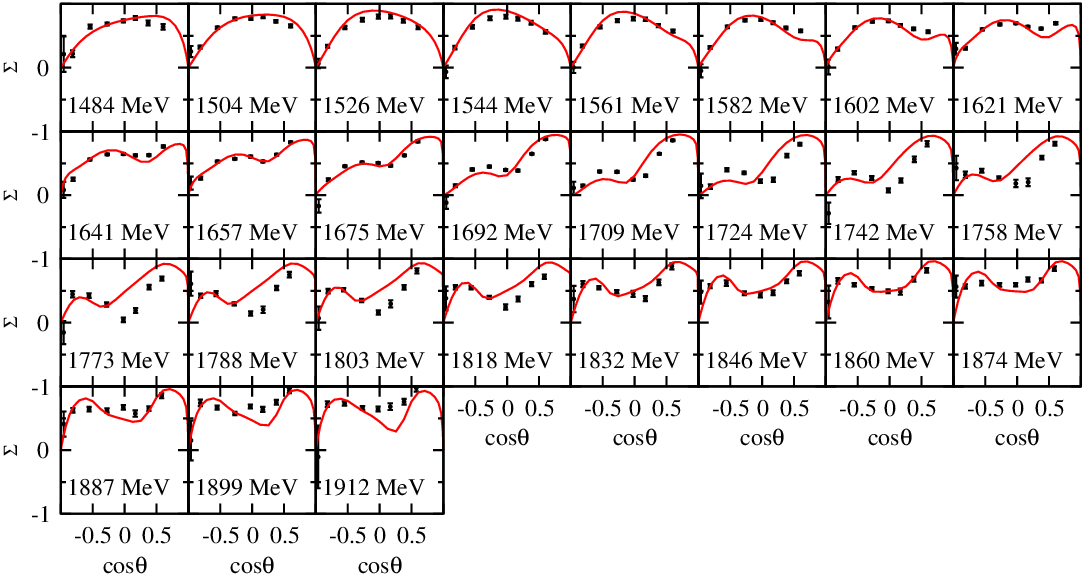}
\caption{\label{fig:gnpi0n-s} Fitted result for $\Sigma$ of $\gamma\, `n\textrm{'} \to \pi^0 n$.}
\end{figure}

Our fits to the $\gamma\, `n\textrm{'} \to \pi^0 n$ data are shown in
Fig.~\ref{fig:gnpi0n-dc} for $d\sigma/d\Omega$ and 
Fig.~\ref{fig:gnpi0n-s} for $\Sigma$.
Here we see the data for $d\sigma/d\Omega$ are very limited,
while reasonably good fits to the $\Sigma$ data have been achieved.
The $\gamma\, `n\textrm{'} \to \pi^0 n$
data are usually extracted from the $\gamma d \to \pi^0 pn$ reaction.
However, the $\gamma d \to \pi^0 pn$ process is more complicated than $\gamma d \to \pi^- pp$
because the final-state interactions between the outgoing $pn$ pair are 
expected to be more sizable, as demonstrated, e.g., in the study~\cite{wsl15} of
the $\gamma d\to \pi NN$ and $\nu d \to l \pi NN$ reactions 
in the $\Delta(1232)$ resonance region.
To improve the fits to the data with $\pi^0 n$ final state, 
it would be more essential to directly analyze the $\gamma d$ reactions 
within our DCC approach in a fully consistent way,
rather than using the $\gamma \, `n\textrm{'} \to \pi N$ data 
extracted by other experiment/analysis groups.

\subsection{Fits to the data for the proton-target reactions}
\label{subsec:p}

\begin{figure}
\includegraphics[clip,width=0.45\textwidth]{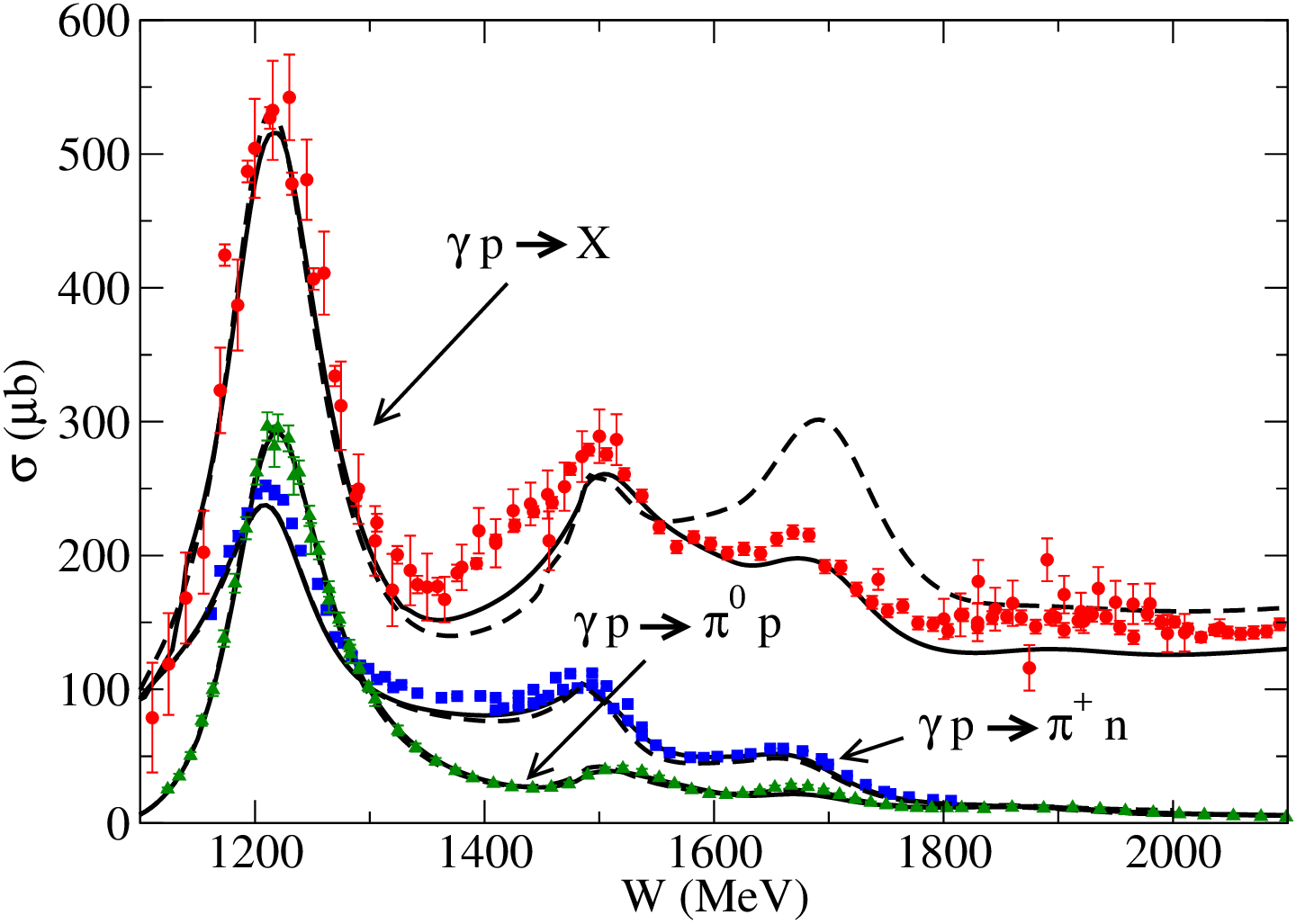}
\caption{(Color online) 
Total cross sections for the inclusive $\gamma p \to X$ and $\gamma p \to \pi N$ reactions.
Solid curves are the results from the current analysis, while 
the dashed curves are the ones obtained from 
the 2013 model~\cite{knls13}.
}
\label{fig:gptot}
\end{figure}

\begin{figure}
\includegraphics[clip,width=0.4\textwidth]{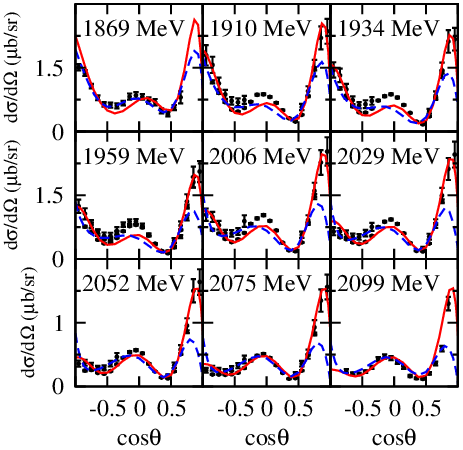}
\qquad 
\includegraphics[clip,width=0.4\textwidth]{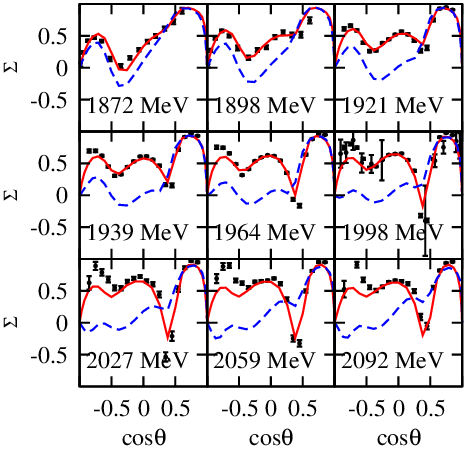}
\caption{(Color online)
$d\sigma/d\Omega$ (left panels) and $\Sigma$ (right panels) of $\gamma p \to \pi^0 p$.
Red solid curves (blue dashed curves) are the result from the current analysis
(the 2013 model~\cite{knls13}).
}
\label{fig:gppi0p}
\end{figure}

\begin{figure}
\includegraphics[clip,width=0.4\textwidth]{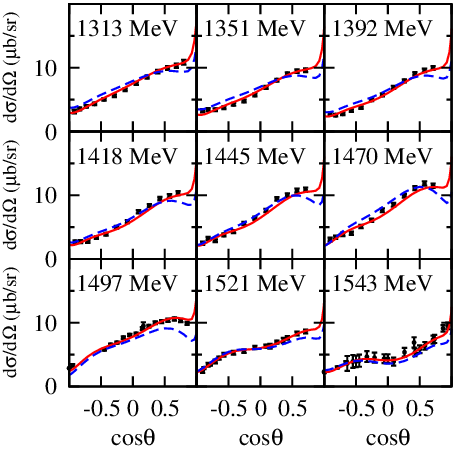}
\qquad
\includegraphics[clip,width=0.4\textwidth]{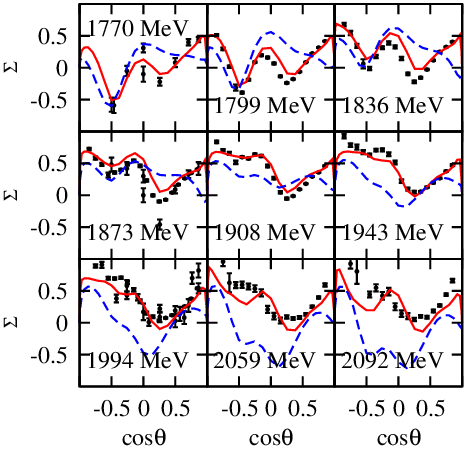}
\caption{\label{fig:gppipn-dc}(Color online)
$d\sigma/d\Omega$ (left panels) and $\Sigma$ (right panels) of $\gamma p \to \pi^+ n$.
Red solid curves (blue dashed curves) are the result from the current analysis
(our previous model~\cite{knls13}).
}
\label{fig:gppipn}
\end{figure}

Since the amount of the data for the proton-target reactions are 
much more than those for the neutron-target reactions, 
our model parameters determined in this work 
are still mainly constrained by the proton-target reactions,
except for the ones associated with the $\gamma n \to N^*$ transitions.
Accordingly, the quality of the fits to the data for the proton-target reactions is similar to
the results we presented in Ref.~\cite{knls13} 
(hereafter we refer to the model of Ref.~\cite{knls13} as ``the 2013 model''). 
Nevertheless, there are some differences that are worth mentioning:
\begin{enumerate}  
\item  
Figure~\ref{fig:gptot} shows the total cross sections for the $\gamma p \to \pi N$ and
inclusive $\gamma p \to X$ reactions, where the solid and dashed curves
are given by the DCC model of this work and the 2013 model~\cite{knls13}, respectively.
We see that both models give almost the same cross sections for $\gamma p \to \pi N$.
However, it turns out that the $\gamma p \to X$ total cross section 
predicted with the 2013 model~\cite{knls13}
visibly overestimates (underestimates) the data at $1.55 \lesssim W \lesssim 1.8$ GeV 
($1.4 \lesssim W \lesssim 1.5$ GeV).
This deviation from the data is mostly due to the uncertainty 
of the $\gamma N \to \pi \Delta, \sigma N, \rho N \to \pi \pi N$ processes
because at present we have included only the
$\gamma p \to \pi N, \eta N, K\Lambda, K\Sigma$ data in our analysis
and the $\pi \pi N$ production processes are indirectly constrained 
through the coupled-channels effect.
To eliminate the deviation from the data mentioned above,
in this work we have included the data for $\gamma p \to X$ total cross section 
at $W \lesssim 1.75$ GeV into our analysis.
As a result, the overestimation of the $\gamma p \to X$ total cross section
at $1.55 \lesssim W \lesssim 1.8$ GeV has been improved significantly in our current model. 
However, the underestimation at $1.4 \lesssim W \lesssim 1.5$ GeV still remains
although some improvements are observed.
To resolve this underestimation, we would need to extend our analysis by including
the $\gamma N \to \pi \pi N$ data and 
by adding some new non-resonant mechanisms for $\gamma N \to \pi \pi N$.
This requires tremendous efforts and time-consuming numerical tasks,
and thus we leave it for our future work.

\item  
In Fig.~\ref{fig:gppi0p}, the results of the fits from this work and the 2013 model~\cite{knls13}
are presented for $d\sigma/d\Omega$ and $\Sigma$ of $\gamma p \to \pi^0 p$.
Overall, a significant improvement has been achieved for these observables in the
high $W$ region with $W\gtrsim 1.87$ GeV.
We further find that the improvements for $d\sigma/d\Omega$ are made mostly 
in the forward angle region, in which the $t$-channel vector-meson exchange processes 
dominate the cross sections and thus mainly contribute to the improvement.
On the other hand, the fitted results for $\Sigma$ are improved in almost 
the entire range of $\cos\theta$, which means that
model parameters associated with bare $N^*$ states are also modified significantly.
As a result, values of the $\gamma p \to N^*$ transition amplitudes
for high-mass $N^*$ resonances are significantly modified as well,
as will be discussed in Sec.~\ref{sec:respar}.

\item 
In Fig.~\ref{fig:gppipn}, the results of the fits from this work and the 2013 model~\cite{knls13}
are presented for $d\sigma/d\Omega$ ($1.4\lesssim W \lesssim 1.5$ GeV) and 
$\Sigma$ ($1.77\lesssim W \lesssim 2.1$ GeV) of $\gamma p \to \pi^+ n$.
Similarly to the $\gamma p \to \pi^0 p$ case,
the results for $d\sigma/d\Omega$ are improved particularly in the forward region, 
while the improvement for $\Sigma$ extends over the entire $\cos\theta$ region.

\end{enumerate}

The fits to the data of the production of $K\Lambda$ and $K\Sigma$ off the proton target
have also been improved significantly both for photon- and pion-induced reactions.
This is, however, not much relevant to the purpose of this paper 
and will not be presented here.

\section{Extracted resonance parameters}
\label{sec:respar}

\begin{table}
\caption{\label{tab:pole} 
Comparison of $N^\ast$ pole mass ($M_R$) and $\pi N$ elastic residue ($R_{\pi N, \pi N}$)
between this work and the 2013 model~\cite{knls13}.
$M_R$ is listed as $\left(\mathrm{Re}(M_R),-\mathrm{Im}(M_R)\right)$ in units of MeV,
while $R_{\pi N,\pi N}=|R_{\pi N,\pi N}|e^{i\phi}$ is listed as $\left(|R_{\pi N,\pi N}|,\phi\right)$ in units of MeV for 
$|R_{\pi N,\pi N}|$ and degree for $\phi$.
The range of $\phi$ is taken to be $-180^\circ \leq \phi < 180^\circ$.
The $N^\ast$ resonances for which the asterisk (*) is marked locate in the complex energy plane
slightly off the sheet closest to the physical real energy axis, yet are still expected
to visibly affect the physical observables. 
}
\begin{ruledtabular}
\begin{tabular}{lccccc}
&                 &\multicolumn{2}{c}{$M_R$}&\multicolumn{2}{c}{$R_{\pi N,\pi N}$} \\
\cline{3-4}
\cline{5-6}
&$J^P(L_{2I2J})$  &This work & 2013 model& This work&2013 model \\
\hline
$N$ baryons
&$1/2^-(S_{11})$ &(1490, 102)& (1482,~~98)*&($  70$,  $-$42)&($ 63$, $ -44$)\\ 
&                &(1652,  71)& (1656,~~85) &($  45$,  $-$74)&($ 53$, $ -70$)\\ 
&$1/2^+(P_{11})$ &(1376,  75)& (1374,~~76) &($  38$,  $-$70)&($ 37$, $ -69$)\\ 
&                &(1741, 139)& (1746,~177) &($  15$,     80)&($ 20$, $   3$)\\ 
&$3/2^+(P_{13})$ &(1708,  65)& (1703,~~70) &($   9$,   $-$4)&($  8$, $  -3$)\\ 
&                &(1765, 160)& (1763,~159) &($  30$, $-$105)&($ 29$, $-106$)\\ 
&$3/2^-(D_{13})$ &(1509,  48)& (1501,~~39) &($  30$,   $-$10)&($ 26$, $ -11$)\\ 
&                &(1702, 148)*&(1702,~141)*&($ <1 $, $-$161)&($  2$, $ 104$)\\ 
&$5/2^-(D_{15})$ &(1651,  68)& (1650,~~75) &($  26$,  $-$27)&($ 28$, $ -31$)\\ 
&$5/2^+(F_{15})$ &(1665,  52)& (1665,~~49) &($  36$,  $-$22)&($ 34$, $ -20$)\\ 
\\                                                   
$\Delta$ baryons                                     
&$1/2^-(S_{31})$ &(1597,  69)&(1592,~~68)  &($  21$, $-$111)&($ 20$, $-111$)\\ 
&                &(1713, 187)&(1702,~193)  &($  20$,     73)&($ 19$, $  65$)\\ 
&$1/2^+(P_{31})$ &(1857, 145)&(1854,~184)  &($  11$, $-$118)&($ 23$, $-123$)\\ 
&$3/2^+(P_{33})$ &(1212,  52)&(1211,~~51)  &($  55$,  $-$47)&($ 53$, $ -47$)\\ 
&                &(1733, 162)&(1734,~176)  &($  16$, $-$108)&($  8$, $-118$)\\ 
&$3/2^-(D_{33})$ &(1577, 113)&(1592,~122)  &($  13$,  $-$67)&($ 18$, $ -62$)\\ 
&                &			-&(1707,~170)* &         	-  &($ 11$, $  49$)\\
&$5/2^-(D_{35})$ &(1911, 130)&(1936,~105)  &($   4$,  $-$30)&($  2$, $ -32$)\\ 
&$5/2^+(F_{35})$ &(1767,  88)&(1765,~~94)  &($  11$,  $-$61)&($ 11$, $ -62$)\\ 
&$7/2^+(F_{37})$ &(1885, 102)&(1872,~103)  &($  49$,  $-$30)&($ 46$, $ -35$)\\ 
\end{tabular}
\end{ruledtabular}
\end{table}

We now turn to discuss the extracted 
$N^*$ resonance parameters, 
which are defined at the poles of the scattering amplitudes in the complex $W$ plane.
In Table~\ref{tab:pole}, we list the pole mass ($M_R$) and $\pi N$
elastic residue ($R_{\pi N,\pi N}$) for each extracted $N^*$ resonance.
Here the corresponding results of the 2013 model~\cite{knls13} 
are also presented for a comparison.
Overall, no significant difference is found in the results between the two models,
implying that pole positions and coupling strengths to the $\pi N$ channel 
are more or less well determined for $N^*$ resonances below $W=2$ GeV.
However, some difference can also be seen 
for the $N^*$ resonances with large imaginary parts for $M_R$,
such as the $P_{31}$ and $D_{35}$ resonances and the second $P_{11}$, $P_{33}$, $D_{33}$ resonances.
This would reflect the fact that analysis dependence comes more into
$N^*$ resonances located far from the physical real energy axis 
and their parameters are in practice less well determined, 
although in principle the pole parameters should be unique.
In particular, we do not find the second $D_{33}$ resonance in the current work,
suggesting that the pole of this resonance has disappeared or moved far away
from the complex-$W$ region close to the physical real energy axis.
One can see qualitatively that there are some correlation in the change of values 
between $\textrm{Im}(M_R)$ and $R_{\pi N,\pi N}$, i.e.,
a larger change in $\textrm{Im}(M_R)$ leads to  a larger change in 
 $R_{\pi N,\pi N}$.
This is perhaps related to the fact that
 the contribution of a resonance to the $\pi N$ partial-wave amplitude
is roughly determined by  the ratio $-iR_{\pi N, \pi N}/\textrm{Im}(M_R)$ at $W\sim\textrm{Re}(M_R)$.
Since the $\pi N$ partial-wave amplitudes are well determined and have almost no difference
between the current and 2013 models,
the values of the $R_{\pi N,\pi N}$ residues tend to vary to ``compensate''
 the change in the extracted values of $\textrm{Im}(M_R)$ such that
the total $\pi N$ partial-wave amplitudes remain the same.

\begin{table}
\caption{\label{tab:helicity_p}
Comparison of helicity amplitudes for the $\gamma p\to N^*$ transition
obtained from this work and the 2013 model~\cite{knls13}.
The presented values follow the notation in Ref.~\cite{ani13}, i.e.,
$A_{1/2,3/2} = \bar A_{1/2,3/2}\times e^{i\phi}$ with $\phi$ taken to be in the range $-90^\circ \leq \phi < 90^\circ$.
The units of $\bar A_{1/2,3/2}$ and $\phi$ are $10^{-3}\ {\rm GeV}^{-1/2}$ and degree, respectively.
Each resonance is specified by the isospin and spin-parity quantum numbers 
as well as the real part of the resonance pole mass.
}
\begin{ruledtabular}
\begin{tabular}{lrrrrrrrr}
& \multicolumn{4}{c}{$A_{1/2}$} & \multicolumn{4}{c}{$A_{3/2}$}  
\\
& \multicolumn{2}{c}{This work}
& \multicolumn{2}{c}{2013 model}
& \multicolumn{2}{c}{This work}
& \multicolumn{2}{c}{2013 model}
\\
\cline{2-3}\cline{4-5} \cline{6-7}\cline{8-9}
Particle $J^P(L_{2I2J})$
&$\bar A_{1/2}$&$\phi$ &$\bar A_{1/2}$&$\phi$ &$\bar A_{3/2}$&$\phi$ &$\bar A_{3/2}$&$\phi$
\\
\hline
$N(1490) 1/2^-(S_{11})$    &$  160 $&$   8  $&$ 161 $&$   9 $&   -    &    -   &-      &-      \\ 
$N(1652) 1/2^-(S_{11})$    &$   36 $&$ -28  $&$  40 $&$ -44 $&   -    &    -   &-      &-      \\ 
$N(1376) 1/2^+(P_{11})$    &$ - 40 $&$  -8  $&$ -50 $&$ -12 $&   -    &    -   &-      &-      \\ 
$N(1741) 1/2^+(P_{11})$    &$ - 47 $&$ -24  $&$  86 $&$ -74 $&   -    &    -   &-      &-      \\ 
$N(1708) 3/2^+(P_{13})$    &$  131 $&$   7  $&$ 234 $&$   2 $&$ - 33 $&$   12 $&$ -70 $&$  -7 $\\ 
$N(1765) 3/2^+(P_{13})$    &$  123 $&$- 11  $&$ 145 $&$ -30 $&$ - 71 $&$    3 $&$ -44 $&$  -1 $\\ 
$N(1509) 3/2^-(D_{13})$    &$ - 28 $&$<  1  $&$ -38 $&$   3 $&$  102 $&$    4 $&$  94 $&$   7 $\\ 
$N(1703) 3/2^-(D_{13})$    &$   13 $&$  50  $&$  26 $&$  64 $&$   31 $&$ - 71 $&$  54 $&$ -42 $\\ 
$N(1651) 5/2^-(D_{15})$    &$    8 $&$  19  $&$   5 $&$ -22 $&$   49 $&$ - 12 $&$  33 $&$ -23 $\\ 
$N(1665) 5/2^+(F_{15})$    &$ - 44 $&$ -11  $&$ -53 $&$  -5 $&$   60 $&$ -  2 $&$  38 $&$   3 $\\ 
\\                                                              
$\Delta(1597)1/2^-(S_{31})$&$  105 $&$   1  $&$ 113 $&$  -1 $&  -     &   -    &-      &-      \\  	 
$\Delta(1713)1/2^-(S_{31})$&$   40 $&$  13  $&$  35 $&$   5 $&  -     &   -    &-      &-      \\  	 
$\Delta(1857)1/2^+(P_{31})$&$  -1 $&$  -78  $&$  52 $&$ -10 $&  -     &   -    &-      &-      \\  	 
$\Delta(1212)3/2^+(P_{33})$&$ -134 $&$ -16  $&$-133 $&$ -15 $&$- 257 $&$  -3  $&$-257 $&$  -3 $\\  	 
$\Delta(1733)3/2^+(P_{33})$&$ - 48 $&$  63  $&$ -72 $&$  71 $&$ - 94 $&$   74 $&$-136 $&$  82 $\\  	 
$\Delta(1577)3/2^-(D_{33})$&$  128 $&$  19  $&$ 129 $&$  17 $&$  119 $&$   46 $&$ 117 $&$  41 $\\  	 
$\Delta(1911)5/2^-(D_{35})$&$   48 $&$ - 22 $&$  53 $&$ -21 $&$   11 $&$  -36 $&$  35 $&$ -15 $\\  	 
$\Delta(1767)5/2^+(F_{35})$&$   38 $&$  - 7 $&$   8 $&$   83$&$  -24 $&$  -80 $&$ -18 $&$ -90 $\\  	 
$\Delta(1885)7/2^+(F_{37})$&$ - 69 $&$  -14 $&$ -62 $&$  -9 $&$ - 83 $&$    2 $&$ -76 $&$   2 $\\  	 
\end{tabular}
\end{ruledtabular}
\end{table}

In Table~\ref{tab:helicity_p}, we present a comparison of the helicity amplitudes 
for the $\gamma p \to N^*$ transition evaluated at the resonance pole positions.
The notation of the presented values follows the one used in Ref.~\cite{ani13}.
In contrast to the $\pi N$ elastic residue $R_{\pi N, \pi N}$,
visible differences from the 2013 model are observed for most of the helicity amplitudes,
except for the very well-established resonances such as the first $P_{33}$ and $S_{11}$ resonances.
This is due to the significant improvement in the fits
to the $\gamma p$ reaction data in this work,  as seen in Figs.~\ref{fig:gptot}-\ref{fig:gppipn}.
In particular, it is found that the change in the helicity amplitudes for $N^*$ resonances with 
the mass $M_R\sim 1.7$ GeV, in particular for  $N(1708)3/2^+$, originates mostly from reducing
the overestimation of the $\gamma p \to X$ total cross sections in our previous 2013 model 
(Fig.~\ref{fig:gptot}).  We also observe that
the improvement in fitting the polarization observables for
$\gamma p \to \pi N, \eta N, K\Lambda, K\Sigma$ reactions at $W \gtrsim 1.8$ GeV
(the right panels of Figs.~\ref{fig:gppi0p} and~\ref{fig:gppipn}
for the case of $\Sigma$ for $\gamma p \to \pi N$) is related to the changes in
the helicity amplitudes of the  higher $N^*$ resonances. 
It is worth mentioning that the $N(1651)5/2^-$ resonance has
small $\bar A_{1/2}$, while the value of $\bar A_{3/2}$ is rather large.
In a constituent quark model, this $N^*$ state is assigned as a member of
the $[{\bf 70}, {^4}8]$ representation of SU(6)$\times$O(3),
and thus the $\gamma p \to N^*$ transition amplitudes are exactly zero~\cite{moorhouse}.
However, we find that within the current DCC model
the large nonzero value of $A_{3/2}$ for the $N(1651)5/2^-$ resonance mostly comes from
the bare $N^*$ contribution. 
This seems to be in contradiction with the above argument based on the naive quark-model 
and a discussion made in Ref.~\cite{ab15}.

One can see from Table~\ref{tab:helicity_p} that the $\gamma p \to N^*$ 
transition amplitudes defined by poles are essentially complex.
Thus they are different from the helicity amplitudes listed by Particle Data Group~\cite{pdg}, 
which are  from the fits by using  the 
Breit-Wigner parametrization and the resulting values are real by definition.
According to a resonance theory based on the Gamow vectors~(see, e.g., Ref.~\cite{madrid}), 
the transition amplitudes defined with the residues at poles of the scattering amplitudes 
are transition matrix elements 
associated with the exact complex-energy eigenstates of the full Hamiltonian of
the considered system obtained under the purely outgoing boundary condition.
Thus the transition amplitudes defined by poles have a clear connection to
the underlying theory, i.e., QCD, while the phenomenological Breit-Wigner parameters do not. 
If the phase $\phi$ of the transition amplitudes is small,
the Breit-Wigner amplitudes can be a good approximation of the pole amplitudes.
In fact, our DCC model give 
$\bar A_{1/2}=-0.133$ GeV$^{-1/2}$ and $\bar A_{3/2}=-0.257$ GeV$^{-1/2}$ for the first $P_{33}$ resonance,
while the (real) Breit-Wigner amplitudes are $A_{1/2}=-0.135\pm 0.006$ GeV$^{-1/2}$
and $A_{1/2}=-0.255\pm 0.005$ GeV$^{-1/2}$~\cite{pdg}. 
{However, if $\phi$ is large, there exists no clear relation between them.}
This argument is of course applicable also to the $\gamma n \to N^*$ transition amplitudes.

\begin{table}
\caption{\label{tab:helicity_n}
Comparison of our helicity amplitudes for the $\gamma n\to N^*$ transition 
with the ones extracted by Bonn-Gatchina (BoGa) analysis~\cite{ani13}.
See the caption of Table~\ref{tab:helicity_p} for the notation of the table.
No resonances corresponding to $N(1765) 3/2^+$ and $N(1703) 3/2^-$ are found in BoGa.
}
\begin{ruledtabular}
\begin{tabular}{lrrrrrrrr}
& \multicolumn{4}{c}{$A_{1/2}$} & \multicolumn{4}{c}{$A_{3/2}$} 
\\
& \multicolumn{2}{c}{This work} & \multicolumn{2}{c}{BoGa} 
& \multicolumn{2}{c}{This work} & \multicolumn{2}{c}{BoGa}
\\
\cline{2-3}
\cline{4-5}
\cline{6-7}
\cline{8-9}
Particle $J^P(L_{2I2J})$ 
&$\bar A_{1/2}$ &$\phi$& $\bar A_{1/2}$ &$\phi$& $\bar A_{3/2}$ &$\phi$& $\bar A_{3/2}$ &$\phi$
\\
\hline
$N(1490) 1/2^-(S_{11})$ &$-112 $&$  16 $&$-103 \pm 11 $&$   8 \pm  5 $&-      &-      &-            &-            \\ 
$N(1652) 1/2^-(S_{11})$ &$-  1 $&$- 47 $&$  25 \pm 20 $&$   0 \pm 15 $&-      &-      &-            &-            \\ 
$N(1376) 1/2^+(P_{11})$ &$  95 $&$- 15 $&$  35 \pm 12 $&$  25 \pm 25 $&-      &-      &-            &-            \\ 
$N(1741) 1/2^+(P_{11})$ &$ 195 $&$-  8 $&$- 40 \pm 20 $&$- 30 \pm 25 $&-      &-      &-            &-            \\ 
$N(1708) 3/2^+(P_{13})$ &$- 59 $&$   6 $&$- 80 \pm 50 $&$- 20 \pm 30 $&$- 28 $&$- 19 $&$-140 \pm 65$&$   5 \pm 30$\\ 
$N(1765) 3/2^+(P_{13})$ &$- 34 $&$  -5 $&-             &-             &$  40 $&$   6 $&-            &-            \\ 
$N(1509) 3/2^-(D_{13})$ &$- 43 $&$-  1 $&$- 49 \pm  8 $&$-  3 \pm  8 $&$-110 $&$   5 $&$-114 \pm 12$&$   1 \pm  3$\\ 
$N(1703) 3/2^-(D_{13})$ &$- 40 $&$ -46 $&-             &-             &$- 77 $&$ -57 $&-            &-            \\ 
$N(1651) 5/2^-(D_{15})$ &$- 76 $&$   3 $&$- 61 \pm  7 $&$- 10 \pm  5 $&$- 38 $&$-  4 $&$- 89 \pm 10$&$- 17 \pm  7$\\ 
$N(1665) 5/2^+(F_{15})$ &$  34 $&$- 12 $&$  33 \pm  6 $&$- 12 \pm  9 $&$- 56 $&$-  4 $&$- 44 \pm  9$&$   8 \pm 10$\\ 
\end{tabular}
\end{ruledtabular}
\end{table}

\begin{table}
\caption{\label{tab:helicity_n-iso}
The isovector and isoscalar 
helicity amplitudes for $\gamma N\to N^*$,
 as defined by Eqs.~(\ref{eq:hel-iso1}) and (\ref{eq:hel-iso0}).
See the caption of Table~\ref{tab:helicity_p} for the notation of the table.
}
\begin{ruledtabular}
\begin{tabular}{lrrrrrrrr}
& \multicolumn{2}{c}{$A^{T=1}_{1/2}$} & \multicolumn{2}{c}{$A^{T=0}_{1/2}$} 
& \multicolumn{2}{c}{$A^{T=1}_{3/2}$} & \multicolumn{2}{c}{$A^{T=0}_{3/2}$} 
\\
Particle $J^P(L_{2I2J})$
&$\bar A^{T=1}_{1/2}$&$\phi$
&$\bar A^{T=0}_{1/2}$&$\phi$
&$\bar A^{T=1}_{3/2}$&$\phi$
&$\bar A^{T=0}_{3/2}$&$\phi$
\\
\hline
$N(1490) 1/2^-(S_{11})$ &$ 136 $&$  11 $&$  26 $&$- 10 $&   -   &   -   &    -  &    -  \\ 
$N(1652) 1/2^-(S_{11})$ &$  19 $&$ -29 $&$  18 $&$- 28 $&   -   &   -   &    -  &    -  \\ 
$N(1376) 1/2^+(P_{11})$ &$- 68 $&$ -13 $&$  28 $&$- 21 $&   -   &   -   &    -  &    -  \\ 
$N(1741) 1/2^+(P_{11})$ &$-120 $&$ -11 $&$  75 $&$-  3 $&   -   &   -   &    -  &    -  \\ 
$N(1708) 3/2^+(P_{13})$ &$  95 $&$   7 $&$  36 $&$   8 $&$  -9 $&$  68 $&$ -29 $&$  -2 $\\ 
$N(1765) 3/2^+(P_{13})$ &$  78 $&$ -10 $&$  45 $&$ -14 $&$ -55 $&$  4  $&$ -16 $&$  -2 $\\ 
$N(1509) 3/2^-(D_{13})$ &$   7 $&$  -2 $&$- 35 $&$  -1 $&$ 106 $&$ 4   $&$- 5  $&$   8 $\\ 
$N(1703) 3/2^-(D_{13})$ &$  20 $&$ -28 $&$ -22 $&$ -63 $&$  54 $&$ -61 $&$ -24 $&$ -48 $\\ 
$N(1651) 5/2^-(D_{15})$ &$  42 $&$   4 $&$- 34 $&$   1 $&$  44 $&$ -8  $&$   6 $&$ -37 $\\ 
$N(1665) 5/2^+(F_{15})$ &$ -39 $&$ -11 $&$  -5 $&$  -8 $&$  58 $&$- 3  $&$   2 $&$  18 $\\ 
\end{tabular}
\end{ruledtabular}
\end{table}

The extracted $\gamma n \to N^*$ transition amplitudes are
listed in Table \ref{tab:helicity_n}. 
Here only the results for the isospin $I=1/2$ nucleon resonances are presented
because $\gamma p \to \Delta^*$ and $\gamma n \to \Delta^*$ give the same value.
In the same table, we also present the results 
obtained by the Bonn-Gatchina (BoGa) analysis~\cite{ani13}
for a comparison.
The results from ours and BoGa show a reasonable agreement for the transition amplitudes,
for which the BoGa analysis assigns relatively small uncertainties for the extracted amplitudes.
In particular, our results for the first $S_{11}$, $D_{13}$, and $F_{15}$ resonances, which
correspond respectively to $N(1535)1/2^-$, $N(1520)3/2^-$, and $N(1680)5/2^+$ in the PDG notation,
are in good agreement with the BoGa results.
However, some disagreement is also seen for several $N^*$ resonances.
The BoGa analysis gives positive $\bar A_{1/2}$ for the second $S_{11}$ resonance,
while in our analysis $\bar A_{1/2}$ is negative and very small.
To obtain a more conclusive result for this transition amplitude, however, we would also need 
to take into account $\eta n$ photoproduction data, as discussed in Ref.~\cite{ani13}.
We hope to make this extended analysis by directly
analyzing $\gamma d \to \eta p n$ reactions, rather than analyzing
the $\gamma\, `n\textrm{'} \to \eta n$ data provided by other experiment/analysis groups,
and this will be presented elsewhere.
The origin of a significant disagreement in the transition amplitudes for the $P_{11}$ resonances
would also come from a couple of reasons:
(a) The pole mass of the second $P_{11}$ resonance from two analyses is different;
i.e., $M_R = 1741-i139$ MeV from our analysis and $M_R = 1687 -i100$ MeV from BoGa.
Since the value of the residue has a strong correlation with
the pole mass and more sensitive to the analysis model used in each analysis,
one should first determine the pole mass well to accomplish a precise determination
of the residues.
(b) The $P_{11}$ resonances are found to give a small contribution 
to the $\gamma n \to \pi N$ reactions within our DCC model, 
and thus the $\gamma n \to N^*$ transition amplitudes are not well constrained 
by the $\gamma n \to \pi N$ data.
Therefore, to get more convergent results, we would need to further include the data 
associated with the other meson photoproductions off the neutron, 
such as $\pi\pi$, $\eta$, and $K$ productions, into our fits to get convergent results.
However, as already mentioned, this is beyond the scope of the current work and will be 
performed elsewhere.

Combining the $\gamma n \to N^*$ and $\gamma p \to N^*$ transition amplitudes
listed in Tables~\ref{tab:helicity_p} and~\ref{tab:helicity_n},
we can obtain the isovector ($T=1$) and isoscalar ($T=0$) parts of
the $\gamma N\to N^*$ amplitude for $I$=1/2 $N^*$
by using the following well-known relations:
\begin{eqnarray}
A^{T=1}_\lambda &=& (A_\lambda^{1/2p}-A_\lambda^{1/2n})/2,
\label{eq:hel-iso1}
\\
A^{T=0}_\lambda &=& (A_\lambda^{1/2p}+A_\lambda^{1/2n})/2,
\label{eq:hel-iso0}
\end{eqnarray}
where 
$A_\lambda^{1/2p}$ and $A_\lambda^{1/2n}$ are 
helicity amplitudes for $\gamma p\to N^*$ and $\gamma n\to N^*$
transitions, respectively.
The resulting values of the isospin-decomposed transition amplitudes within our current DCC analysis 
is presented in Table~\ref{tab:helicity_n-iso} as a reference.

\section{Summary and discussions}
\label{sec:summary}

In this work, we have extended our DCC analysis~\cite{knls13} of
the $\pi p, \gamma p \to \pi N, \eta N, K\Lambda, K\Sigma$ reactions
by further including the data for pion photoproductions off the neutron target, 
$\gamma n \to \pi N$, in the fits.
The helicity amplitudes for the $\gamma n \to N^*$ transition, 
 defined by the residues of the poles of the scattering amplitudes, have then been extracted.
Through this combined analysis of both the proton- and neutron-target reactions,
the resonance pole masses and the $\gamma p \to N^*$ transition amplitudes
extracted from our previous analysis~\cite{knls13}
have also been revised accordingly. 
Our results allow an isospin decomposition of the
$\gamma N \rightarrow N^*$ transitions, which is needed
for testing the hadron structure calculations
and investigating the neutrino-induced reactions~\cite{nks15}.
The extracted $\gamma n \to N^*$ transition amplitudes are compared  with
the results of the BoGa analysis.
It is found that  two results are consistent
with each other overall.
However, some significant disagreements also exist for several $N^*$ resonances,
implying that further extensions of both  analyses to analyze more complete data on
the neutron target will be needed to make progress.

As mentioned throughout this paper,
in our current analysis we have used the $\gamma\, `n\textrm{'} \to \pi N$ data 
extracted from the $\gamma d \to \pi NN$ data by other experiment/analysis groups.
In most experimental analyses, the $\gamma n \to \pi N$ cross sections and polarization observables
are extracted by simply applying momentum cuts to the deuteron data and choosing
the kinematics where the quasi-free mechanisms are assumed to dominate the reaction processes.
Effects of the nucleon Fermi motion inside the deuteron have also been 
included in some analyses, but final $\pi NN$ interactions are usually neglected.
On the other hand, several theoretical investigations~\cite{darwish,fix,lev06,sch10,wsl15}
have shown that the $\pi NN$ final-state interaction  
has very large effects on the $\gamma d\to\pi^0 pn$ reaction.
Similar large effects are expected also for the  other neutral-meson productions such as 
$\gamma d \to M^0 pn$ with $M^0 = \eta,\eta',\omega,\phi,\cdots$.
To make further progress in the study of the $N^*$ spectroscopy,
an  approach for investigating  these meson production reactions off the deuteron 
must be developed. 
Since the accuracy of the  extracted  $\gamma n \to \pi N$ data 
depends on the way of unfolding the many body effects from the raw data,
it is highly desirable to analyze directly the data of $\gamma d \to \pi NN$ 
reactions based on a well-developed reaction model.
The DCC model employed in our analysis is particularly useful for the 
analysis of $\gamma$-d and also $\gamma$-nuclei reactions since 
the necessary off-shell amplitudes are readily available.
To complete such a  analysis, a method to describe the $\pi NN$ dynamics
in the $\Delta$ and higher $N^*$ resonance region has to be explored.
These necessary tasks towards determining electromagnetic interactions
associated with the $N^*$ resonances will be taken step by step and
presented elsewhere.

\begin{acknowledgments}
This work was supported by the Japan Society for the Promotion of Science (JSPS) 
KAKENHI Grant No.~25800149 (H.K.) and 
No.~16K05354 (T.S.), the MEXT KAKENHI Grant No.~25105010 (T.S.),
and by the U.S. Department of Energy, Office of Nuclear Physics Division, 
under Contract No. DE-AC02-06CH11357.
This research used resources of the National Energy Research Scientific Computing Center,
which is supported by the Office of Science of the U.S. Department of Energy
under Contract No. DE-AC02-05CH11231, and resources provided on Blues and/or Fusion,
high-performance computing cluster operated by the Laboratory Computing Resource Center
at Argonne National Laboratory. 
\end{acknowledgments}

\end{document}